\documentclass[10pt,twocolumn]{article}%
\usepackage{latex8}
\usepackage{times}
\usepackage{amsfonts}
\usepackage{amsmath}
\usepackage{amssymb}
\usepackage{graphicx}%
\begin{document}

\title{Towards a Real-Time Data Driven Wildland Fire Model}
\author{Jan Mandel\\Department of Mathematical Sciences\\University of Colorado Denver\\Denver, CO 80217-3364\\Jan.Mandel@gmail.com
\and Jonathan D. Beezley\\Department of Mathematical Sciences\\University of Colorado Denver\\Denver, CO 80217-3364\\jon.beezley.math@gmail.com
\and Soham Chakraborty\\Department of Computer Science\\University of Kentucky\\Lexington, KY 40506-0046\\sohaminator@gmail.com
\and Janice L. Coen\\Mesoscale and Microscale Meteorology Division\\National Center for Atmospheric Research\\Boulder, CO 80307-3000\\janicec@ucar.edu
\and Craig C. Douglas\\Department of Computer Science\\University of Kentucky\\Lexington, KY 40506-0046\\craig.c.douglas@gmail.com
\and Anthony Vodacek\\Center for Imaging Science\\Rochester Institute of Technology\\Rochester, NY 14623\\vodacek@cis.rit.edu
\and Zhen Wang\\Center for Imaging Science\\Rochester Institute of Technology\\Rochester, NY 14623\\zxw7546@cis.rit.edu
}
\maketitle

\begin{abstract}
A wildland fire model based on semi-empirical relations for the spread rate of a surface
fire 
and post-frontal heat release is coupled with
the Weather Research and Forecasting atmospheric model (WRF).  The propagation of
the fire front
is implemented by a level set method.
Data is assimilated by a morphing ensemble Kalman filter, which provides
amplitude as well as position corrections. Thermal images of a fire will provide
the observations and will be compared to a synthetic image from the model state.
\end{abstract}

\thispagestyle{empty}


\section{Introduction}

In this paper, we review the current state of a project in wildland fire
simulation based on real-time sensor monitoring.  The model will be driven by
real-time data and run on remote supercomputers. The state of the model will be adjusted from real-time
airborne imagery as well as ground sensor data.
We have also used a regularization approach to EnKF for wildfire
\cite{Johns-2007-CEK-short} with a fire model by reaction-diffusion-convection
partial differential equations \cite{Mandel-2006-WMD}.
See \cite{Mandel-2007-DDD-short} for an earlier picture of this project.


\section{Coupled fire-atmosphere model}
\label{sec:coupled-model}

This section is based on \cite{Mandel-2007-DAW}, with some new developments.
Our current code implements the mathematical ideas from \cite{Clark-2004-DCA}
in a simplified contemporary numerical framework in that the fire propagation,
originally done by a custom ad hoc tracer code, is now done using a simpler level set
method, and the fire model is now coupled with the Weather Research and
Forecasting model (WRF, www.wrf-model.org), a community supported numerical
weather prediction code.
This adds capabilities to the codes used in previous work such as
theoretical experiments of \cite{Clark-2004-DCA} and
\cite{Coen-2005-SBE}, where it was used in reanalysis of a real fire.
Specifically, it allows taking
advantage of the WRF architecture, which provides a natural parallelization of
both the fire and the atmospheric code in shared as well as distributed memory
and support for data assimilation.


\begin{figure}[t!]
\begin{center}
\includegraphics[width=3.5in]{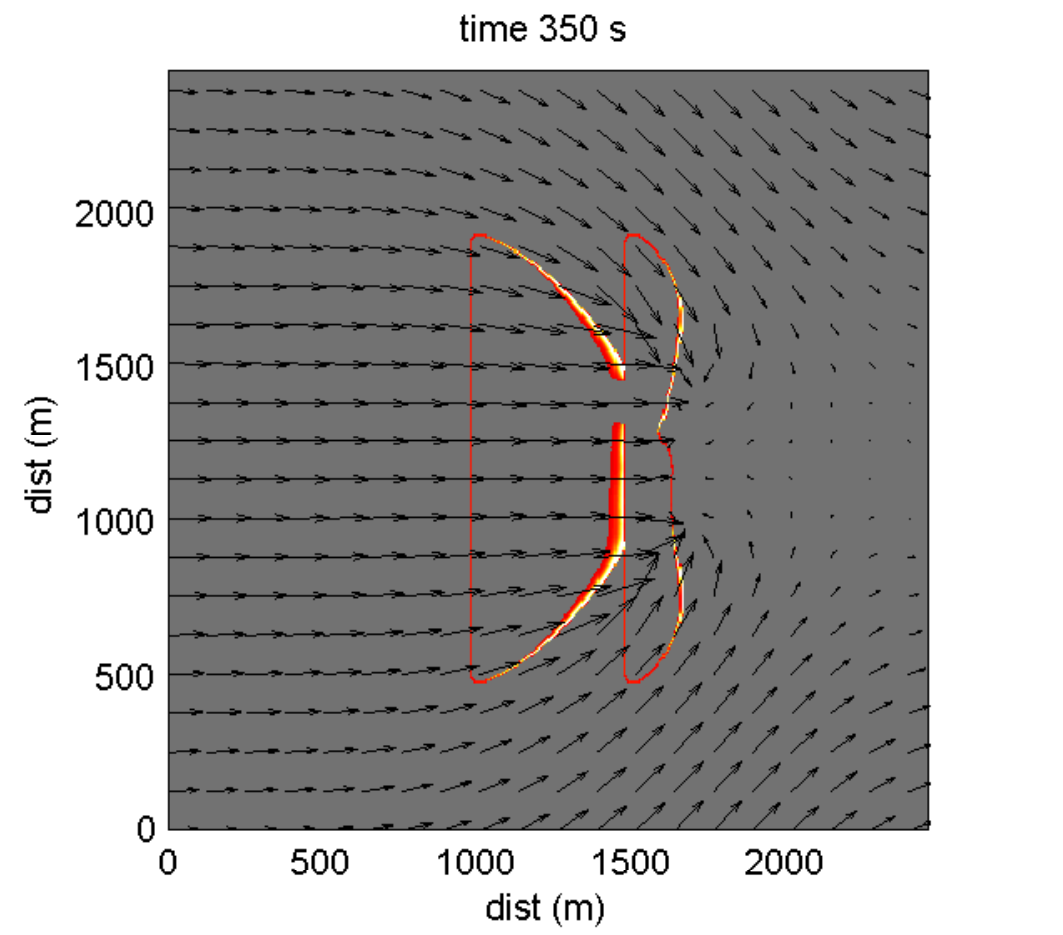}
\end{center}
\par
\caption{Coupled fire-atmosphere simulation. Fire propagates from two line
ignitions and one circle ignition, which are in the process of merging. The
arrows are horizontal wind at ground level. False color is fire heat flux.
The fire front on the right has
irregular shape and is slowed down because of air being pulled up by the heat
created by the fire. This kind of fire behavior cannot be modeled by empirical
spread models alone.}%
\label{fig:prop}%
\end{figure}

\subsection{Fire spread model}
\label{sec:fire-spread}

We use a semi-empirical fire propagation model
\cite{Clark-2004-DCA,Rothermel-1972-MMP} to represent a fire spreading on the surface.
Given wind speed $\overrightarrow{v}$
and terrain height $z$, the model postulates that the fireline
evolves with the spread rate $S$ in the normal direction to the fireline
$\overrightarrow{n}$ given by
$$
S=R_{0}+a\left(  \overrightarrow{v}\cdot\overrightarrow{n}\right)^b+
d\nabla z\cdot\overrightarrow{n},
$$
where $R_{0}$, $a$, $b,$ and $d$ are fuel coefficients
determined from laboratory experiments.  In addition, the spread
rate is cut off to satisfy $0\leq S\leq S_{\max}$, where $S_{\max}$ depends on
the fuel. Based on laboratory experiments, the model further assumes that after ignition the fuel amount decreases as an
exponential function of time, the characteristic time scale depending on the
fuel type (rapid mass loss in grass, slow mass loss in larger fuel particles).
The output of the model is the sensible and the latent heat fluxes (temperature and water vapor)
from the fire to the atmosphere,
taken to be proportional to the amount of fuel burned.


\subsection{Propagation by level set method}

In the level set method, the burning area at time $t$ is represented as
$\left\{  \left(  x,y\right)  :\psi\left(  x,y,t\right)  <0\right\}  $, where
$\psi$ is the \emph{level set function}. The fireline at time $t$ is the level
set $\left\{  \left(  x,y\right)  :\psi\left(  x,y,t\right)  =0\right\}  $.
The normal to the fireline can be computed from the level set function as
$
\overrightarrow{n}=\nabla\psi/\left\Vert \nabla\psi\right\Vert
$.
The level set function satisfies the differential equation
$
{\partial\psi}/{\partial t}+S\left\Vert \nabla\psi\right\Vert =0
$,
which we solve numerically by the Heun's method (i.e., Runge-Kutta method of
order 2) using
approximation of $\nabla\psi$ from upwinded approximation of $\nabla\psi$ by Godunov's method%
: each partial derivative is approximated by the
left difference if both the left and the central differences are nonnegative,
by the right difference if both the right and the central differences are
nonpositive, and taken as zero otherwise. The reason for using Heun's
method is not accuracy but conservation: the explicit Euler method
systematically
overestimates $\psi$ and thus slows down fire propagation or even stops it
altogether while Heun's method behaves reasonably well.
The level set function is
initialized to the signed distance from the fireline.


\subsection{Coupling with weather}
\label{sec:coupling}

The fire model takes as input the horizontal wind velocity components and outputs the heat
fluxes.  The finest atmospheric mesh interfaces with the fire.  However, WRF meshes cannot be
nested vertically, so even the finest mesh still goes vertically over the
whole atmosphere.  We have used time step $0.5$s with the $60$m finest atmospheric
mesh step and $6$m fire mesh step, which satisfied the CFL\ stability
conditions in the fire and in the atmosphere.

The heat flux from the fire model cannot be applied directly as a boundary
condition on the derivatives of the corresponding physical field (air
temperature or water vapor contents) because WRF\ does not support flux
boundary conditions. Instead, the flux is inserted by modifying the
temperature and water vapor concentration over a depth of many cells, with
exponential decay away from the boundary.


\section{Data assimilation}
\label{sec:data-assimilation}

\begin{figure*}[t!]
\begin{center}
\includegraphics[width=5in]{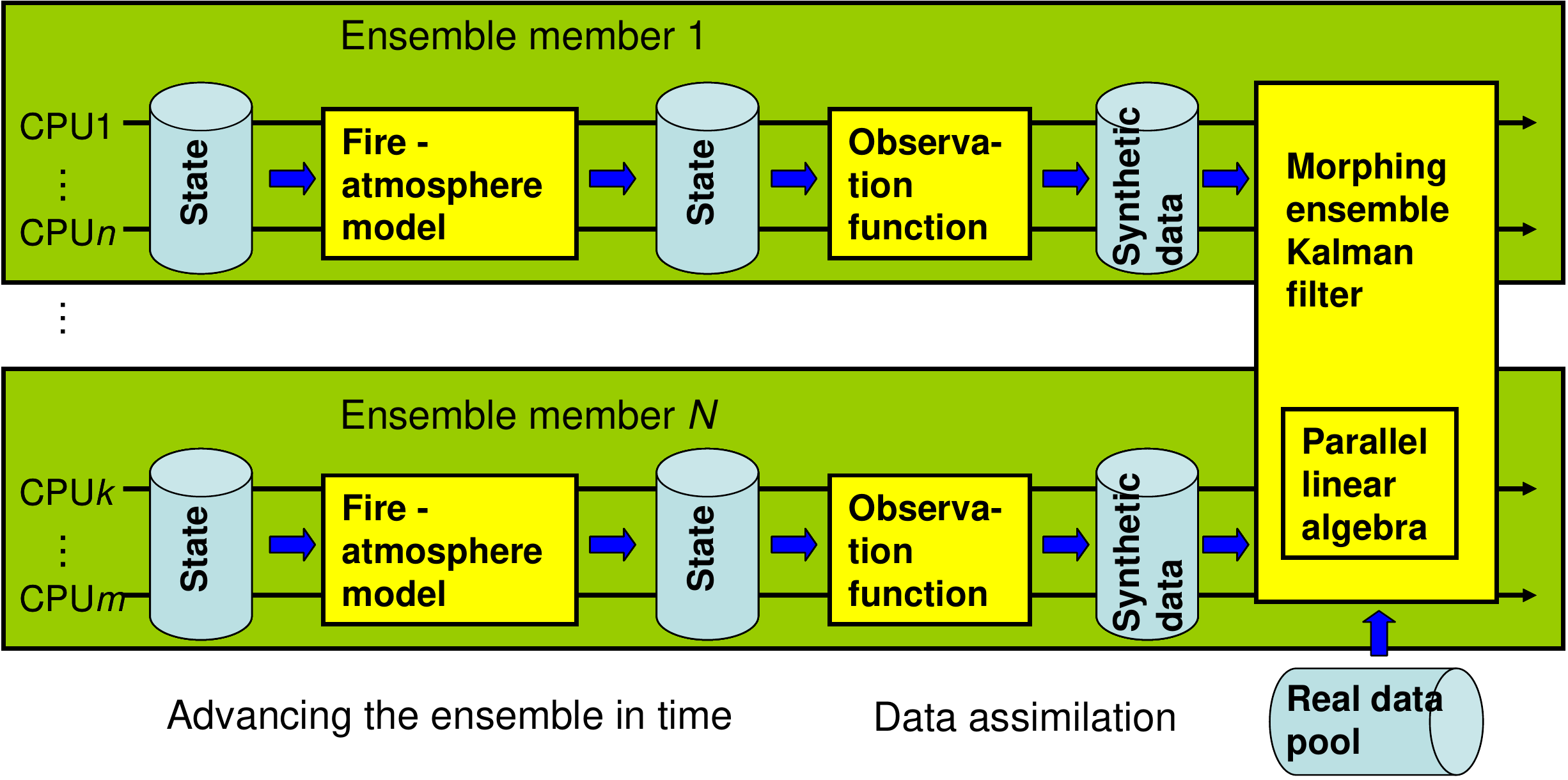}
\end{center}
\par
\caption{Parallel implementation of one assimilation cycle. Ensemble members are advanced in time
and the observation function evaluated for each ensemble member independently on a subset of
processors. The morphing EnKF adjusts the member states by comparing the synthetic
data with real data and balances the uncertainty in the data, which is given as error bound,
with the uncertainty in the simulation, computed from the spread of the whole ensemble. The morphing
EnKF runs on all processors.
The ensemble of model states is maintained in
disk files. The observation function takes as input the disk files
and delivers synthetic data also in disk files. The EnKF inputs the synthetic data
and the real data, and modifies the files with the ensemble states. The model, the
observation function, and the EnKF are in separate executables.
}%
\label{fig:parallel}%
\end{figure*}

The overall scheme is in Fig.~\ref{fig:parallel}.
We are using the ensemble Kalman filter (EnKF) \cite{Evensen-2003-EKF} for data
assimilation.
We currently know
how to work with data from airborne/space images and land-based
sensors, though work remains on making all the connections described.
 In the rest of this section we describe how to treat various kinds of
data (Sec.~\ref{sec:craig}), how can synthetic image data for the
fire be obtained from the system state (Sec.~\ref{sec:tony}), and how
is are the model states adjusted by in response the the data
(Sec.~\ref{sec:enkf}).

\subsection{Data sources}
\label{sec:craig}

To date, our efforts have been to get data from a variety of sources (e.g.,
weather stations, aerial images, etc.) and to compare it
\cite{Chakraborty-2008-DAV} to data from the fire-atmosphere code \cite{Clark-2004-DCA}.
A state vector is handed to the observation function routines and the vector
is modified and returned. The state is transferred using disk files.
Individual subvectors corresponding to the most common variables are
extracted or replaced in the files.
A software layer exists in order to hide both the fire code and the data
transfer method so that the code is not dependent on a particular fire-atmosphere code.

Consider an example of a weather station that reports its location, a
timestamp, temperature, wind velocity, and humidity. The atmosphere code has multiple nested grids.
For a given grid, we have to determine in which cell the weather station is
located, which is done using linear interpolation of the location.
The data is is determined at relevant grid points using biquadratic
interpolation. We compare the computed results with the weather station data.
We determine if a fireline is in the cell (or neighboring ones) with the
weather station temperature to see if there really is a fire in the cell.
Currently, the state vector is updated for the temperature and returned for further
processing. In future, this will be replaced by the output of synthetic data in disk files for the data
assimilation, as in Fig.~\ref{fig:parallel}.

\subsection{Synthetic scene generation}
\label{sec:tony}

The purpose of synthetic scene generation \cite{Wang-2008-MWF} is to use the model state to produce an image
just like one from the infrared camera, so that it can be compared to the actual data
in the data assimilation.
The synthetic scene generation depends on the propagation model,
parameters such as velocity of the fire movement, as well as its output (the heat flux), and
the input (the wind speed).  The use of the parameters from the model allows us to estimate the 3D
flame structure, which provides an effective geometry for simulating
radiance from the fire scene.  Given the 3D flame structure, we assume we can
adequately estimate the infrared scene radiance by including three aspects of
radiated energy.  These are radiation from the hot ground under the fire
front and the cooling of the ground after the fire front passes, which
accounts for the heating and cooling of the 2D surface, the direct
radiation to the sensor from the 3D flame, which accounts for the intense
radiation from the flame itself, and the radiation from the 3D flame that
is reflected from the nearby ground.  This reflected radiation is most
important in the near and mid-wave infrared spectrum.  Reflected long-wave
radiation is much less important because of the high emissivity (low
reflectivity) of burn scar in the long-wave infrared portion of the spectrum
\cite{Kremens-2003-MTT}.

\begin{figure}[t!]
\begin{center}
\includegraphics[width=3.5in]{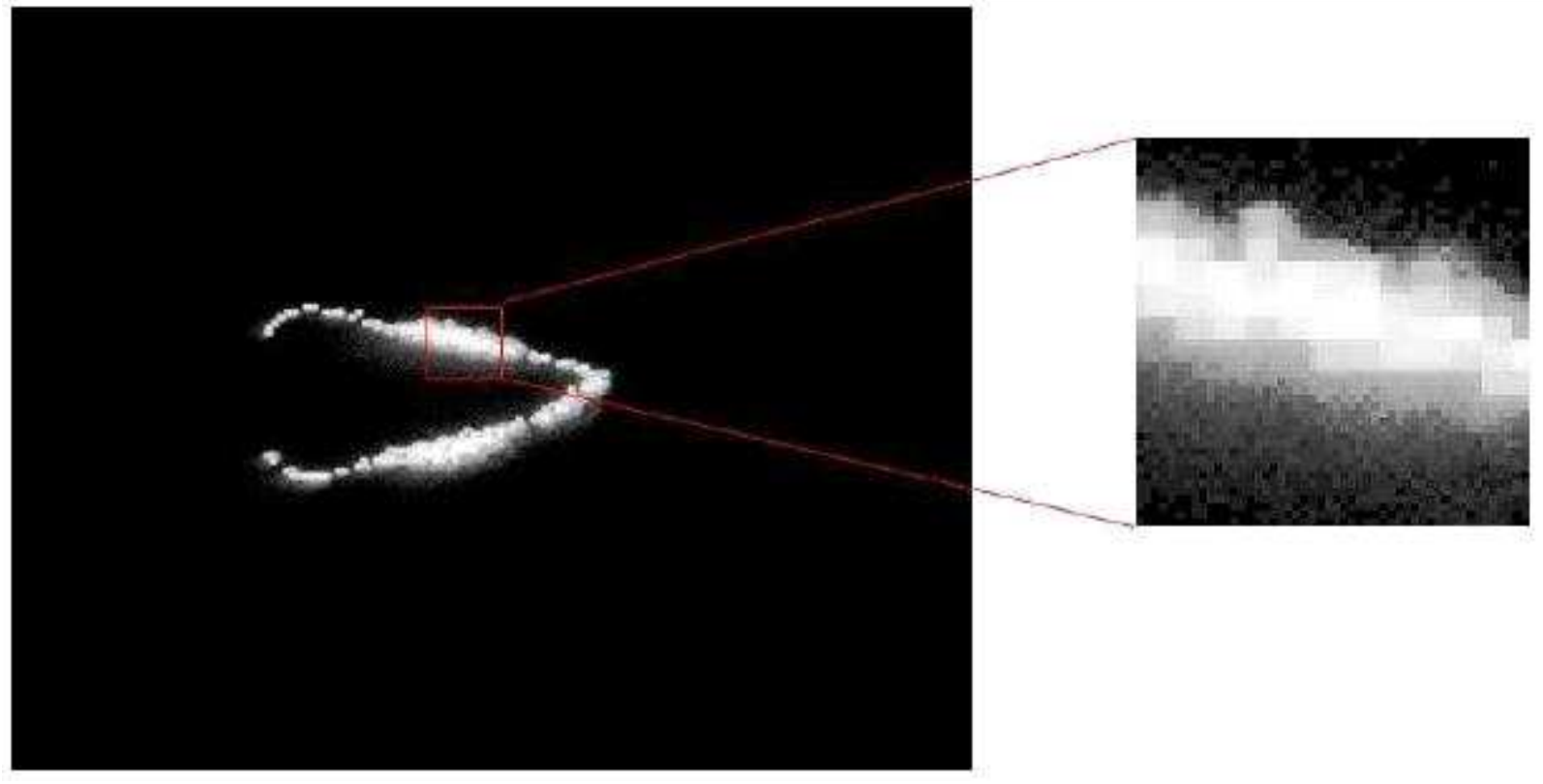}
\end{center}
\par
\caption{Synthetic fire scene.  A DIRSIG rendering of the midwave (3-5 micrometers)
infrared radiation from a modeled grassfire.  The scene is rendered as it would
be observed with RIT's WASP airborne infrared camera system flying about 3000 m
above ground. The zoomed area shows the effect of the hot ground and voxelized
flame structure in the central blocky pixels and the lighter gray fading away
at the edges is the reflected radiation from the 3D flame. From \cite{Wang-2008-MWF}.}%
\label{fig:synfire}%
\end{figure}

The 2D fire front and cooling are estimated with a double exponential.  The
time constants are 75 seconds and 250 seconds and the peak temperature at the
fire front is constrained to 1075K.  The position of the fire front is
determined from the propagation model described above.  The 3D flame
structure is estimated by using the heat release rate and experimental
estimates of flame width and length and the flame is tilted based on wind
speed.  This 3D structure is represented by a 3D grid of voxels.

The 2D ground temperatures and the 3D voxels representing the flame are inputs
to a ray tracing code for determining the radiance from those sources as they
would be viewed by an airborne remote sensing system.  The ray tracing code
used is the Digital Imaging and Remote Sensing Image Generation Model
(DIRSIG).  The Digital Imaging and Remote Sensing Image Generation (DIRSIG)
model is a first principles based synthetic image generation model developed
by the Digital Imaging and Remote Sensing Laboratory at RIT
\cite{DIRSIG-2006,Schott-1999-ASI}.
The model can produce multi- or hyper-spectral imagery from the visible
through the thermal infrared region of the electromagnetic spectrum.  Radiance
reflected from the ground and the effects of the atmosphere are include in the
radiance calculation.  The resulting synthetic radiance image (Fig.~\ref{fig:synfire})
was validated by
calculation of the fire radiated energy and comparing those results to
published values derived from satellite remote sensing data over wildland
fires \cite{Wooster-2003-FRE}.  We are continuing to work on synthetic image
generation with the goal of replacing the computationally intensive, but
accurate, ray tracing method with a simpler method of calculating the fire
radiance based upon the radiance estimations that are inherent in the fire
propagation model.


\subsection{Morphing ensemble Kalman filter}
\label{sec:enkf}

This section again follows \cite{Mandel-2007-DAW}. The state of the model
consists of the level set function $\psi$ and the ignition time $t_{i}$, both
given as arrays of values associated with grid nodes. These grid arrays can be
modified by data assimilation methods with relative ease, unlike the tracers
in \cite{Clark-2004-DCA}. Data assimilation \cite{Kalnay-AMD-2003} maintains
an approximation of the probability distribution of the state. In each
analysis cycle, the probability distribution of the state is advanced in time
and then updated from the data likelihood using the Bayes theorem.
EnKF represents the probability distribution of the state by an
ensemble and it uses the model only as a black box, without any additional
coding required. After advancing the ensemble in time, the EnKF replaces the
ensemble by its linear combinations with the coefficients obtained by solving
a least squares problem to balance the change in the state and the difference
from the data.

However, the EnKF applied directly to the model fields does not work well when
the data indicate a fire in a different location than in the state, because such data have
infinitesimally small likelihood and
the span of the ensemble does contain a state consistent with the data.
Therefore, we adjust also the simulation state by changing the position of the fire
rather than an additive correction alone, by borrowing the techniques of
registration and morphing from image processing. Essentially, we replace the
linear combinations of states in the EnKF by intermediate states obtained by
morphing, which leads to the morphing EnKF method \cite{Beezley-2008-MEK}.

\begin{figure}[t!]
\begin{center}%
\begin{tabular}
[c]{cc}%
\hspace*{-0.15in}\includegraphics[width=1.65in]{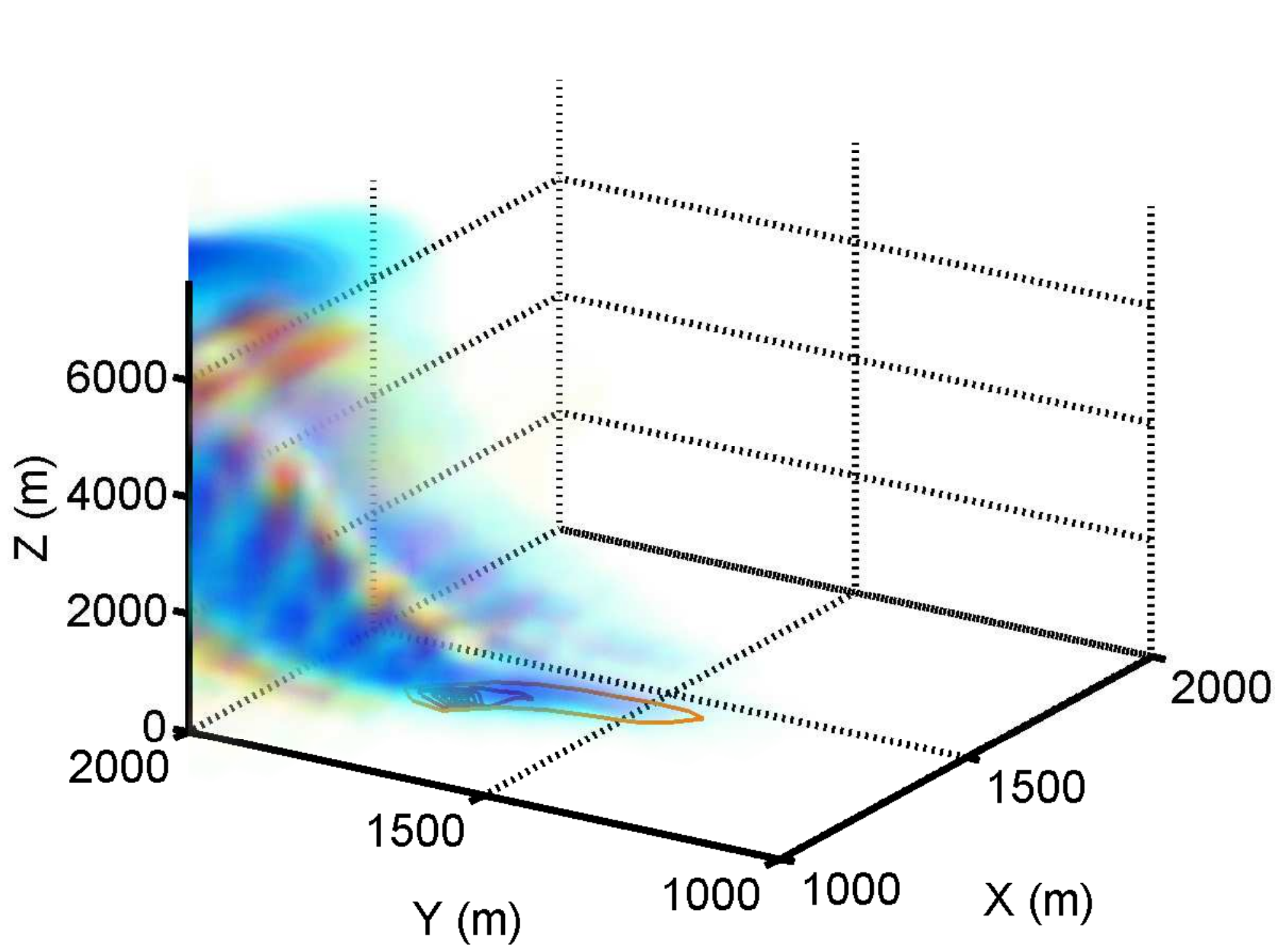} \hspace*{-0.15in.pdf} &
\hspace*{-0.15in} \hspace*{-0.15in}\includegraphics[width=1.65in]{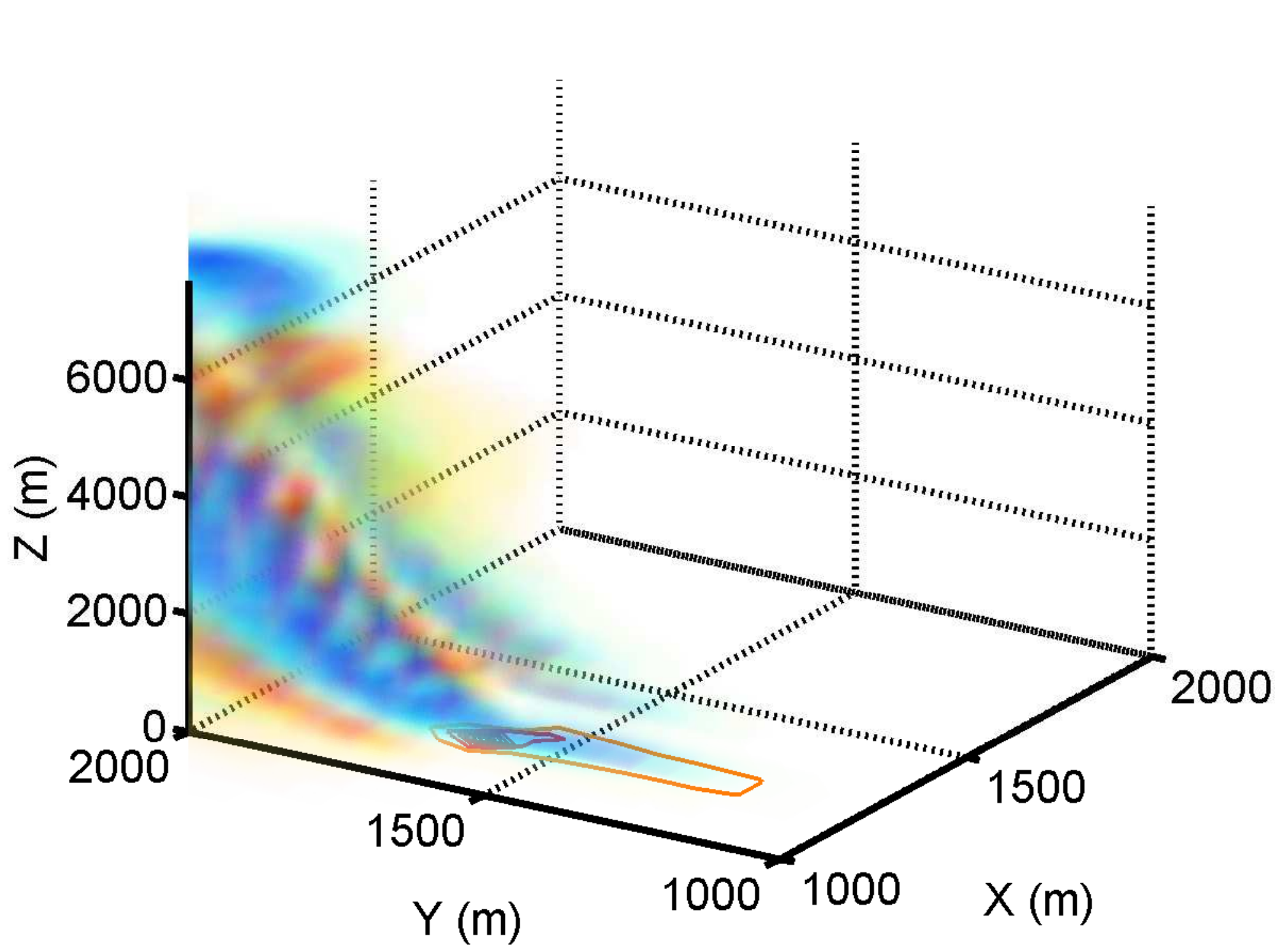}\\
(a) & (b)\\
\hspace*{-0.15in}\includegraphics[width=1.65in]{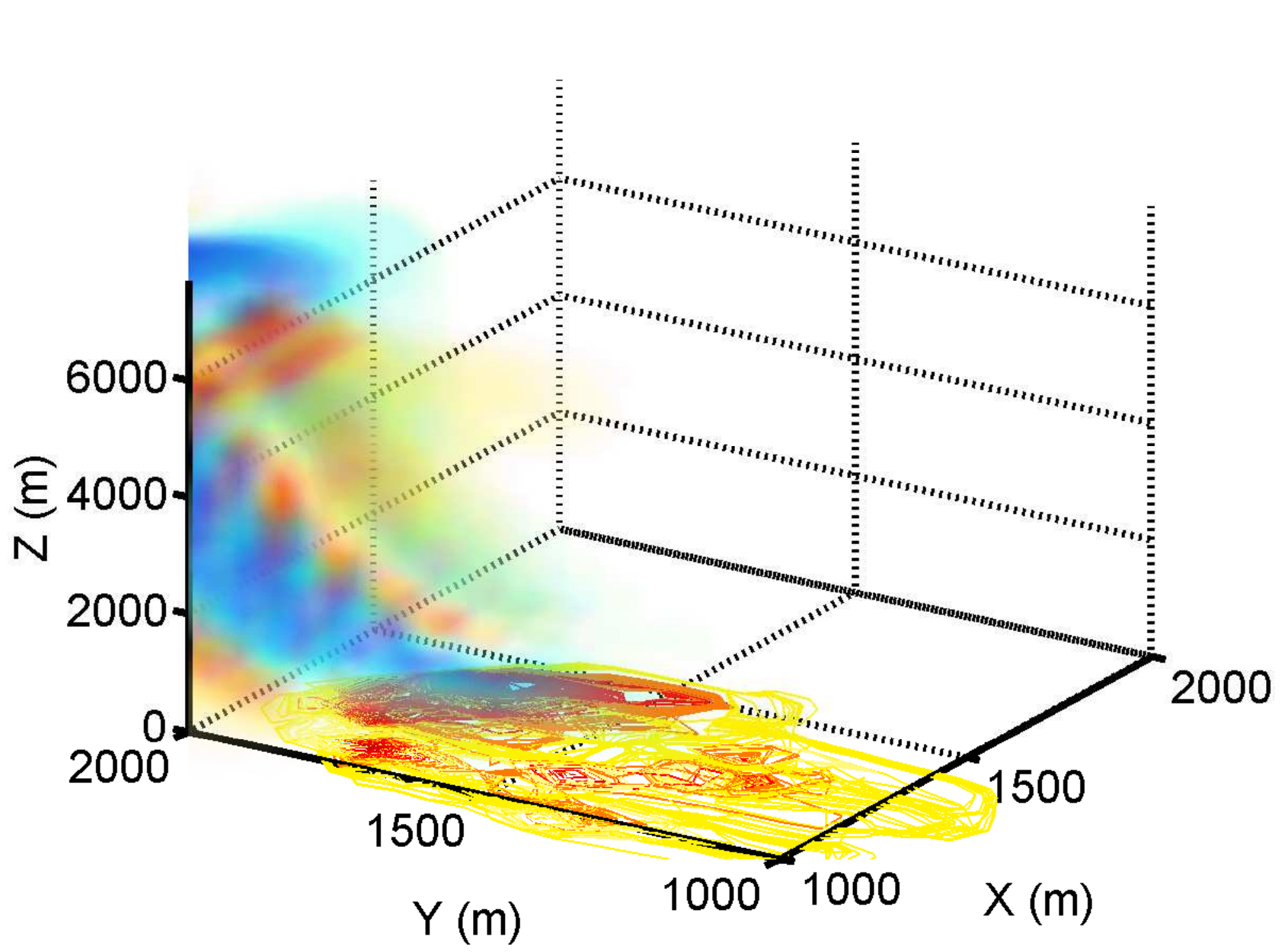} \hspace*{-0.15in.pdf} &
\hspace*{-0.15in} \hspace*{-0.15in}\includegraphics[width=1.65in]{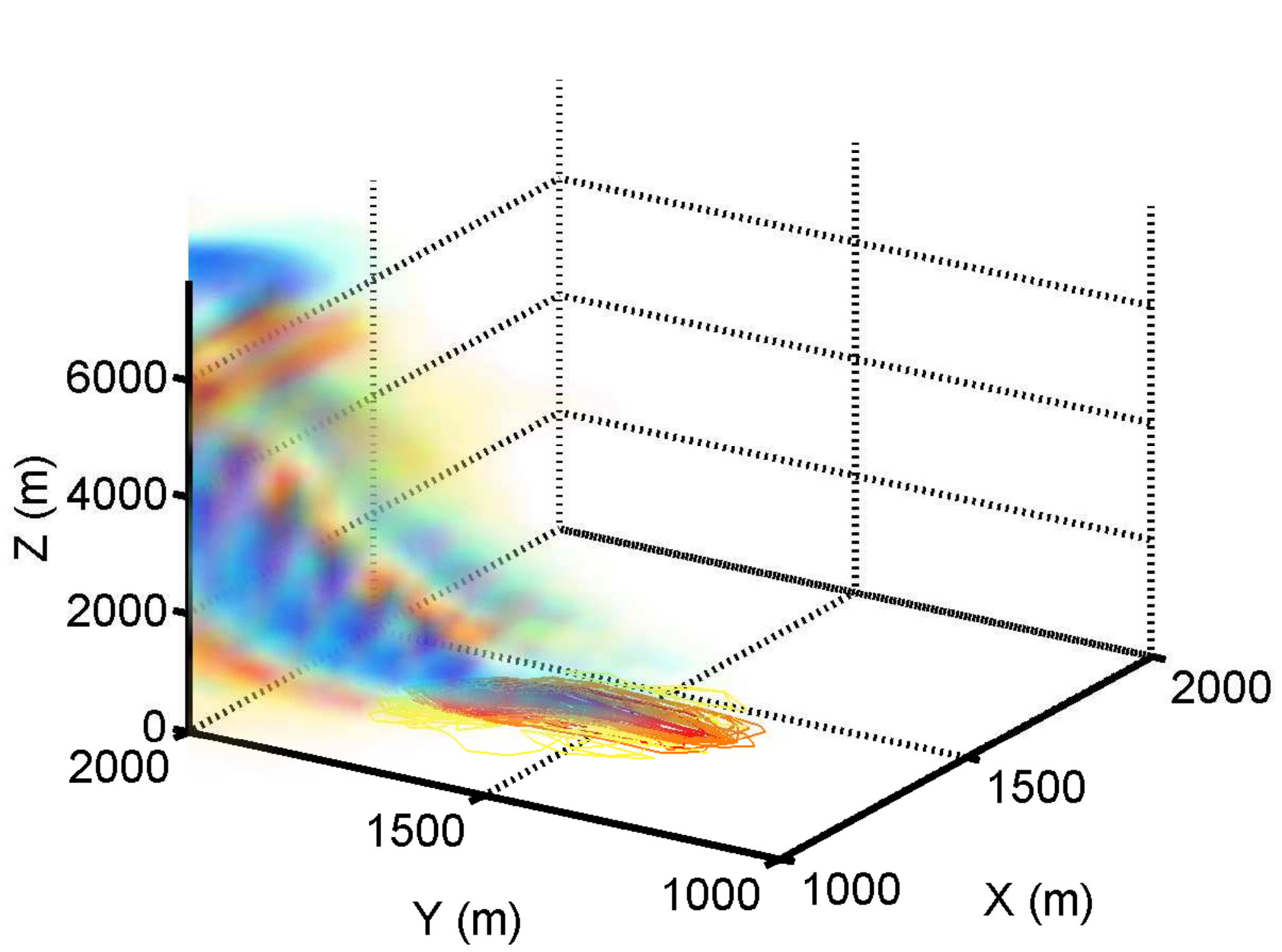}\\
(c) & (d)
\end{tabular}
\newline
\end{center}
\caption{The morphing EnKF applied to the fireline propagation model coupled
with WRF. False color and contour on the horizontal plane is the fire heat
flux. The volume shading is the vorticity of the atmosphere. The reference
solution (a) is the simulated data. The initial ensemble was created by a
random perturbation of the comparison solution (b), with the fire ignited at
an intentionally incorrect location. The standard ENKF (c) and the morphing
EnKF (d) were applied after 15 minutes. The ensembles have 25 members each,
with the heat fluxes shown superimposed. The standard EnKF ensembles diverges
from the data, while the morphing EnKF ensemble keeps closer to the data.
Reproduced from \cite{Mandel-2007-DAW}.}%
\label{fig:morph-coupled}%
\end{figure}

Given two functions $u_{0}$, $u$, representing the same physical field, such
as the temperature, or the level set function, from two states of the coupled
model, registration can be described as finding a mapping $T$ of the spatial
domain so that $u\approx u_{0}\circ\left(  I+T\right)  $, where $\circ$
denotes the composition of mappings and $I$ is the identity mapping. The field
$u$ and the mapping $T$ are given by their values on a grid. To find the
registration mapping \ $T$ automatically, we solve approximately an
optimization problem of the form%
\[
\left\Vert u-u_{0}\circ\left(  I+T\right)  \right\Vert +\left\Vert
T\right\Vert +\left\Vert \nabla T\right\Vert \rightarrow\min.
\]
We then construct intermediate functions $u_{\lambda}$ between $u_{0}$ and
$u_{1}$ by \cite{Beezley-2008-MEK}
\begin{equation}
u_{\lambda}=\left(  u+\lambda r\right)  \circ\left(  I+\lambda T\right)
,\quad0\leq\lambda\leq1, \label{eq:intermediate}%
\end{equation}
where $r=u\circ\left(  I+T\right)  ^{-1}-u_{0}$. The morphing EnKF works by
transforming the ensemble member into extended states of the form $\left[
r,T\right]  $, which are input into the EnKF. The result is then converted
back by (\ref{eq:intermediate}). See Fig. \ref{fig:morph-coupled} for an
illustrative result.

\section{Acknowledgement}

This work was supported by NSF grants
CNS-0325314, CNS-0324910, CNS-0324989, CNS-0324876,
CNS-0540178,
CNS-0719641, CNS-0719626, CNS-0720454,
DMS-0623983,
EIA-0219627,
and
OISE-0405349.


\bibliographystyle{latex8}
\bibliography{../../bibliography/dddas-jm}


\end{document}